\documentclass[sn-mathphys-num, iicol]{sn-jnl}

\usepackage{graphicx}
\usepackage{calrsfs}
\usepackage{xcolor}
\usepackage{amsmath} 

\begin{document}

\title{Ordered and Disordered Skyrmion States on a Square Substrate}

\author*[1]{J. C. Bellizotti Souza}
\email{jc.souza@unesp.br}

\author[2]{C. J. O. Reichhardt}
\author[2]{C. Reichhardt}
\author[3]{N. P. Vizarim}
\author[4]{P. A. Venegas}

\affil*[1]{POSMAT - Programa de P\'os-Gradua\c{c}\~ao em Ci\^encia e Tecnologia de Materiais, S\~ao Paulo State University (UNESP), School of Sciences, Bauru 17033-360, SP, Brazil}

\affil[2]{Theoretical Division and Center for Nonlinear Studies, Los Alamos National Laboratory, Los Alamos, New Mexico 87545, USA}

\affil[3]{``Gleb Wataghin'' Institute of Physics, University of Campinas, 13083-859 Campinas, S\~ao Paulo, Brazil}

\affil[4]{Department of Physics, S\~ao Paulo State University (UNESP), School of Sciences, Bauru 17033-360, SP, Brazil}

\date{\today}

\abstract{
We examine the ordering of skyrmions interacting with a square substrate created from a modulation of anisotropy using atomistic simulations. We consider fillings of {\it f} = 1.5, 2.0, 2.5, 3.0, and 3.5 skyrmions per potential minimum as a function of magnetic field and sample size. For a filling of {\it f} = 2.0, we find various dimer orderings, such as tilted dimer states, as well as antiferromagnetic ordering that is similar to the colloidal dimer ordering seen on square substrates. The ability of the skyrmions to change shape or annihilate produces additional states that do not occur in the colloidal systems. For certain parameters at {\it f} = 2.0, half of the skyrmions can annihilate to form a square lattice, or a superlattice of trimers and monomers containing skyrmions of different sizes can form. At lower fields, ordered stretched skyrmion states can appear, and for zero field, there can be ordered stripe states. For {\it f} = 3.0, we find ferromagnetic ordered trimers, tilted lattices, columnar lattices, and stretched phases. For fillings of {\it f} = 1.5 and 2.5, we find bipartite lattices, different ordered and disordered states, and several extended disordered regions produced by frustration effects.
}

\maketitle

\section{Introduction}

Particles interacting with a smooth background will form a lattice structure that minimizes the particle-particle interaction energy; however, when the particles also interact with a periodic substrate, they can form a variety of ordered, partially ordered, or disordered states due to the competition
between the particle-particle interaction energy and substrate energy \cite{Bak82, Hallen93, Field02, Mikhael08, Reichhardt17}. The
type of order depends on the number of particles per potential minima as well as the symmetry and strength of the substrate. Examples of systems of this type include charged colloids on periodic optical substrates \cite{Brunner02, Bohlein12},
colloids on patterned grooved surfaces \cite{OrtizAmbriz16}, colloids on magnetic substrates \cite{Loehr16}, vortices in type-II superconductors with periodic pinning arrays
\cite{Harada96, Reichhardt98, Field02}, vortices in Bose
Einstein condensates on optical trap arrays
\cite{Tung06}, cold atom systems \cite{Bloch05}, dusty plasmas \cite{Huang22},
and Wigner crystal ordering in mori{\' e} systems \cite{Li24, Reichhardt25}.

Ordering of atoms and molecules on periodic substrates has been
studied for monomers, dimers, or more complex molecular states interacting with square or
triangular substrates \cite{Coppersmith82, Ramseyer98}.
Particle ordering on periodic substrates is typically characterized by
the filling factor
$f = N/N_m$, where $N$ is the number of particles and
$N_m$ is the number of substrate minima.
At integer fillings,
the particles typically form an ordered commensurate lattice
with the same symmetry as the substrate,
while at noninteger fillings, there may be
a commensurate lattice containing well defined
defects such as interstitials or vacancies, a fully disordered state,
or, if the substrate is weak,
a partially ordered or floating state
may appear \cite{Reichhardt17}.
When there are two or more particles
inside each substrate minimum,
the particles can form an effective $n$-mer state that can
have both positional and
orientational ordering \cite{Brunner02, Reichhardt02}.
This type of $n$-mer ordering has been observed
for charged colloids on two-dimensional periodic substrates and
named ``colloidal molecular crystals''
in analogy to molecular crystals where similar ordering appears
\cite{Reichhardt02, Brunner02, Agra04, Sarlah05, Sarlah07, ElShawish08}.

In the colloidal systems for a filling of
$f = 2.0$ on a triangular substrate, possible ordered
dimer states include herringbone arrangements as
well as fully aligned states analogous to ferromagnetic ordering.
When the
colloids are on a square substrate, there is an
effective quadrupole interaction between adjacent dimers,
favoring perpendicular alignment of the dimers and leading to
an antiferromagnetic ordering \cite{Agra04}.
For $f = 3.0$ on a triangular substrate, a trimer state can appear in which
all of the trimers are aligned with each other \cite{Reichhardt02, Brunner02},
while on a square substrate,
the trimers align into columns of alternating orientation
\cite{Reichhardt02}.
When the temperature is increased or
the particle-particle interaction strength
is decreased, colloidal molecular crystals
can undergo multiple step disordering transitions
in which the dimers or trimers first
lose their orientational ordering but remain confined to the substrate
minima, and later begin to hop out of the minima to
form a modulated fluid \cite{Reichhardt02, Brunner02, Mikulis04, Sarlah07}.
For rational fillings such as $f=3/2$ or $5/2$, other types
of ordered or disordered states can occur.
For example, at a filling of $f=3/2$ on a triangular substrate, the
system forms a disordered mixture of monomers and dimers,
while at $f=7/4$ there is
an ordered mixture of monomers and dimers \cite{Reichhardt05}.

Similar molecular crystal type ordering
has been studied for vortices in type-II superconductors
on periodic substrates when there are multiple
vortices per substrate minimum
\cite{Neal07, Reichhardt07}.
More recently, ordered Wigner molecular crystal
states have been directly imaged in a charge ordering system
on a triangular substrates for fillings of
$f = 2$, 3, and higher \cite{Li24}.

In particle based systems such as colloids,
the particle number is conserved and the particles cannot
deform.
It is not known how the ordered and disordered
molecular crystal states on a periodic substrate
might be modified when the particles in the molecular crystals
can distort, change size, or annihilate.
Such deformations are possible
in magnetic skyrmions, which are bubble-like
textures that interact repulsively with each other
to form a triangular lattice in the absence of a substrate
\cite{Muhlbauer09, Yu10, Nagaosa13, EverschorSitte18},
similar to the behavior of colloidal particles, Wigner crystals,
and superconducting vortices.
Skyrmions can change size or shape as a function of
applied magnetic field \cite{EverschorSitte18, Wu21},
and can also change shape when interacting with a substrate \cite{Reichhardt22}.
Skyrmions are promising candidates
for numerous applications including memory devices \cite{Fert13, Sisodia22}
and novel computing approaches \cite{Prychynenko18, Grollier20, Gomes25}.
Since there is considerable interest in creating dense arrays of skyrmions
for memory, understanding how to pack or order the
skyrmions in a controlled
fashion would be very valuable,
and skyrmion ordering on periodic substrates is a promising direction
for producing dense skyrmion assemblies.

Recently, it was proposed that skyrmions on a triangular substrate could form
what are called skyrmion molecular crystals when
there are two or three skyrmions
per substrate minimum \cite{Souza25}. Here
the skyrmions could readily distort or annihilate.
For some parameters, the skyrmions act like
stiff particles and form herringbone or ferromagnetic
dimer states and ferromagnetic trimer states similar to those
found for colloidal systems.
Due to their deformability, the skyrmions can also form states
not found for stiff particles, including
superlattice states containing different skyrmion sizes,
such as large monomers coexisting with small dimers and trimers.
At lower fields, stretched states
can appear in which a portion of the skyrmions are elongated into
merons while
the rest of the skyrmions remain circular.
Situations in which skyrmions could
be made to interact with periodic substrates
include moir\'e systems \cite{Tong18, Hejazi21, Akram21},
nanostructured samples
\cite{Zhang21, Juge21, Reichhardt22},
or skyrmions coupled to periodically ordered
vortices in type-II superconductors \cite{Neto22}.
Skyrmions in liquid crystal systems
could interact with periodic arrays of optical traps \cite{Duzgun20}.

In this work, we examine the ordering of skyrmions on
a square substrate for
fillings of $f = 1.5$, 2.0, 2.5 and $3.0$
as a function of magnetic field and substrate lattice constant.
The square substrate symmetry competes with the
natural triangular symmetry of the skyrmion lattice, and can lead to
effects not found for a triangular substrate
\cite{Souza25}.
For $f = 2.0$, when the skyrmions are small
we find an antiferromagnetic state and a tilted dimer lattice
similar to those observed in colloid systems \cite{Reichhardt02, Agra04}.
At higher fields, a superlattice checkerboard state appears
in which large monomers alternate with aligned trimers,
so that the net filling is still $f = 2.0$.
For other parameters, half of the skyrmions annihilate and a square
skyrmion lattice appears with an $f =1.0$ filling.
At lower fields, the dimers can form a variety of elongated meron states,
but long-range orientational order is preserved.
We also find parameters at which disordered mixtures of
monomers, dimers, and trimers without orientational ordering
appear, which is likely due to a frustration effect.

At a skyrmion filling of $f = 3.0$ on the square substrate,
we find ordered columnar, ferromagnetic,
and tilted states, as well as stretched states containing
elongated merons coexisting
with circular skyrmions.
We observe states in which
all of the skyrmions annihilate, as well as
superlattice states containing skyrmions of different sizes.
When $f = 3/2$, we
find several bipartite lattices consisting of
square commensurate ordering with additional
filled rows in every other column,
as well as states with coexisting monomers and dimers.
We observe several complex stretched phases containing
stripes and skyrmions in which
the system has long-range order.
For $f = 5/2$, we find
ordered trimer-dimer checkerboard lattices
in which all of the trimers have one orientation and all of the dimers
have another orientation.
There are large regions of field and substrate lattice spacing
for which the
$f = 5/2$ system does
not exhibit long-range order.
We show that
even at zero fields,
the periodic substrate can induce long-range order in the
elongated dimer stripe and meron states.

Our work is relevant to a variety of systems
on square substrates
that combine a particle-like nature with the ability to distort,
such as charged drops, emulsions, binary fluids, skyrmions in liquid crystals,
and other magnetic systems.
The orientational order of the
skyrmion trimer or dimer states could be used to store information,
making them possible components for a memory device.

\section{Methods}\label{sec:2}

\begin{figure}
  \centering
  \includegraphics[width=\columnwidth]{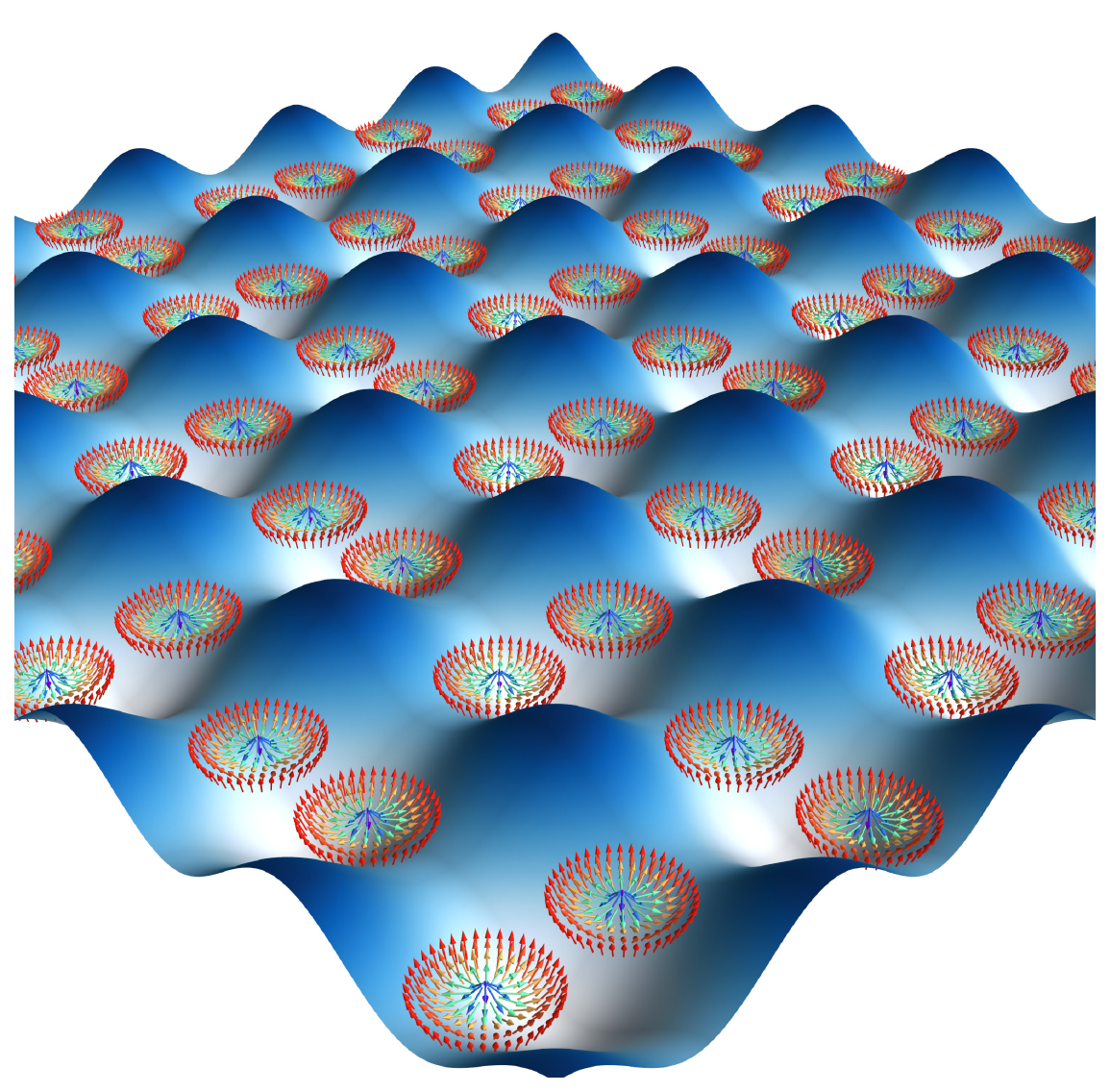}
  \caption{
    A schematic of the system showing the value of the perpendicular
    magnetic anisotropy (PMA) modulated according to $K(x, y)$.
    White areas are PMA minima and blue areas are PMA maxima.
    Colored arrows indicate the magnetic moments inside the skyrmions
    at a filling of $f=2$.
  }
  \label{fig:1}
\end{figure}

We model skyrmions on a square lattice using atomistic simulations,
which take into account the dynamics of individual atomic magnetic
moments \cite{Evans18}. We consider
a ferromagnetic ultrathin film that can sustain N{\' e}el skyrmions.
Our sample is of size $L\times L$ with periodic boundary conditions
along the $x$ and $y$ directions.

The periodic substrate is generated by varying the
sample anisotropy according to
$K(x, y)=\frac{K_0}{4}\left[\cos\left(\frac{2\pi x}{a_0}\right)+\cos\left(\frac{2\pi y}{a_0}\right) + 2\right]$, where
$K_0$ is the anisotropy depth and $a_0=L/6$ is the substrate lattice constant.
The value of $a_0$ is adjusted for samples with different
sizes so that all samples contain $N_m=36$ well-defined minima.
This produces the square anisotropy substrate
depicted schematically in Fig.~\ref{fig:1}.

The Hamiltonian governing the dynamics of the atomistic magnetic moments
is given by
\cite{Evans18, Iwasaki13, Iwasaki13a}:

\begin{eqnarray}\label{eq:1}
  \mathcal{H}=&-\sum_{\langle i, j\rangle}J_{ij}\mathbf{m}_i\cdot\mathbf{m}_j
  -\sum_{\langle i, j\rangle}\mathbf{D}_{ij}\cdot\left(\mathbf{m}_i\times\mathbf{m}_j\right)\nonumber\\
  &-\sum_i\mu\mathbf{H}\cdot\mathbf{m}_i
  -\sum_{i} K(x_i, y_i)\left(\mathbf{m}_i\cdot\hat{\mathbf{z}}\right)^2 \ .
\end{eqnarray}
In our system we apply a magnetic field perpendicular to the sample
along $-z$.
The film is modeled as a square arrangement of atoms with a lattice constant $a=0.5$~nm.
The first term on the right hand side
of Eq.~(1) is the exchange interaction with an exchange constant of
$J_{ij}=J$ between magnetic moments
$i$ and $j$.
The second term is the interfacial Dzyaloshinskii–Moriya
interaction, where $\mathbf{D}_{ij}=D\mathbf{\hat{z}}\times\mathbf{\hat{r}}_{ij}$ is the Dzyaloshinskii–Moriya
vector between magnetic moments $i$ and $j$ and $\mathbf{\hat{r}}_{ij}$
is the unit distance vector between sites $i$ and $j$.
Here, $\langle i, j\rangle$ is the sum over only the
first neighbors of the $i$th magnetic moment.
The Zeeman interaction
with an applied external magnetic field $\mathbf{H}$
is represented by the third term.
The magnitude of the magnetic moment is $\mu=g\mu_B$ where
$g=|g_e|=2.002$ is the electron $g$-factor
and $\mu_B=9.27\times10^{-24}$~J~T$^{-1}$ is the
Bohr magneton.
The last term gives the perpendicular magnetic anisotropy (PMA), where
$x_i$ and $y_i$ are the spatial coordinates of the $i$th
magnetic moment.
In ultrathin films the long-range dipolar interactions act as a
PMA \cite{Wang18} and simply shift the effective PMA values.

The time evolution of atomic magnetic moments is obtained using the
Landau-Lifshitz-Gilbert (LLG)
equation \cite{Seki16, Gilbert04}:

\begin{equation}\label{eq:2}
  \frac{\partial\mathbf{m}_i}{\partial t}=-\gamma\mathbf{m}_i\times\mathbf{H}^\mathrm{eff}_i
  +\alpha\mathbf{m}_i\times\frac{\partial\mathbf{m}_i}{\partial t}
  \ .
\end{equation}
Here $\gamma=1.76\times10^{11}~$T$^{-1}$~s$^{-1}$ is the electron gyromagnetic ratio and
$\mathbf{H}^\mathrm{eff}_i=-\frac{1}{\mu}\frac{\partial \mathcal{H}}{\partial \mathbf{m}_i}$
is the effective magnetic field which includes all the interactions from
the Hamiltonian. The phenomenological Gilbert damping
term is $\alpha$.

The material parameters are $J=1$~meV, $D=0.5J$,
$K_0=0.2J$, and $\alpha=0.3$.
For each simulation, the system is initiated with $f=N/N_m$
skyrmions per anisotropy minimum, where we focus on
$f=3/2$, 2.0, 5/2, and 3.0. An example of an $f=2.0$ state
appears in Fig.~\ref{fig:1}.
The numerical integration
of Eq.~\ref{eq:2} is performed with thermal effects,
$\mathbf{H}^\mathrm{eff}_i=-\frac{1}{\mu}\frac{\partial \mathcal{H}}{\partial {\bf m}_i}+\mathbf{H}^\mathrm{th}$, where $\mathbf{H}^\mathrm{th}=\hat{\boldsymbol{\Gamma}}\sqrt{\frac{2\alpha k_BT}{\gamma\mu\Delta t}}$ is the thermal effect
field, $\hat{\boldsymbol{\Gamma}}$ is a unit vector whose components
follow a Gaussian distribution of random numbers, $k_B$ is the Boltzmann
constant, $T$ is the temperature, and $\Delta t=0.01\hbar/J$ is the integration
time step. The temperature starts at $T=5$ K and linearly decreases over
a time period
of 20 ns until $T=0$ K. The system then relaxes with no thermal effects
for a period of 20 ns. This procedure is
repeated six times per value of $\mu H$
and $L$. The construction of the phase diagrams is performed by inspecting
the lowest energy states.


\section{$f =  2.0$}

\begin{figure}
  \centering
  \includegraphics[width=\columnwidth]{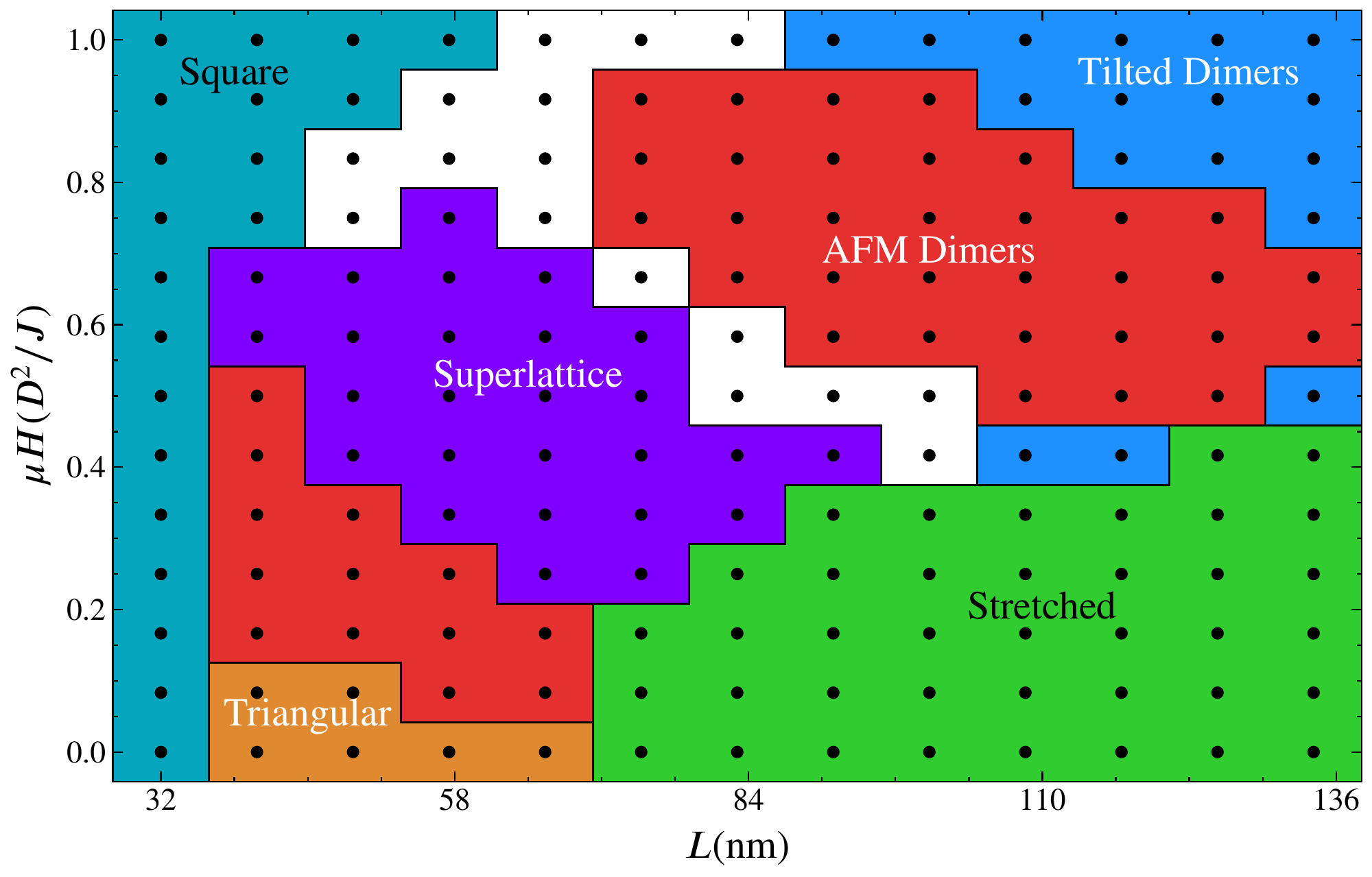}
  \caption{
    Phase diagram indicating the skyrmion ordering as a function of magnetic
    field $\mu H$ vs system size $L$
    for a system with $N_\mathrm{sk}/N_m=2$ as the
    starting condition.
    Teal: square lattice.
    Orange: triangular lattice.
    Purple: monomer-trimer superlattice.
    Red: antiferromagnetic (AFM) dimer state.
    Green: stretched state, where individual skyrmions elongate
    as a consequence of the low magnetic field values.
    Blue: tilted dimers, where each dimer aligns along 45$^\circ$ but
    adjacent dimers have opposite tilt angles.
    In white regions, the system is either disordered or does not have
    a well defined ordering.
  }
  \label{fig:2}
\end{figure}

\begin{figure}
  \centering
  \includegraphics[width=\columnwidth]{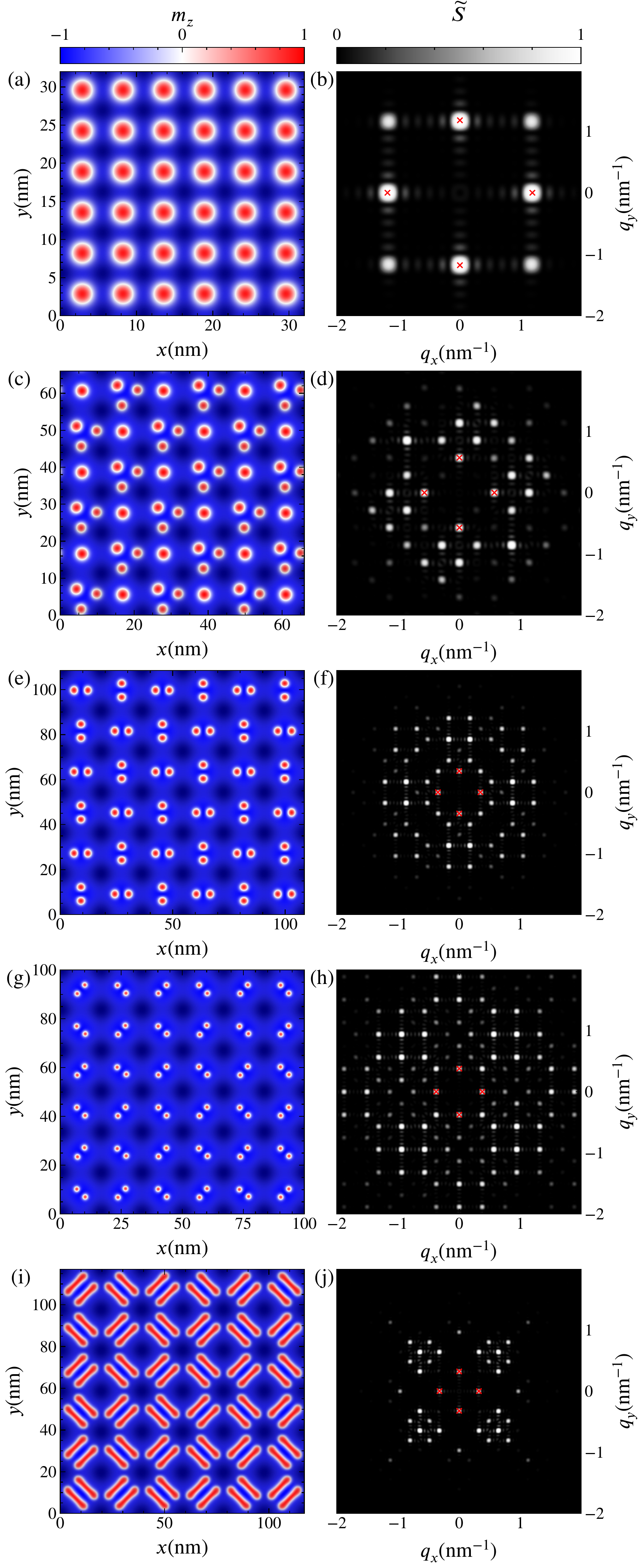}
  \caption{
    Configurations of the $N_\mathrm{sk}/N_m=2$ skyrmion
    lattice for selected states
    from the phase diagram in Fig.~\ref{fig:2}. (a, c, e, g, i) Real
    space images. (b, d, f, h, j) Reciprocal space images.
    (a, b) Square lattice at $\mu H=0.33D^2/J$ and $L=32$ nm.
    (c, d) Superlattice at $\mu H=0.58D^2/J$ and $L=66$ nm.
    (e, f) AFM dimer state at $\mu H=0.58D^2/J$ and $L=108.5$ nm.
    (g, h) Tilted dimer state at $\mu H=1D^2/J$ and $L=100$ nm.
    (i, j) Stretched state at $\mu H=0.25D^2/J$ and $L=117$ nm.
  }
  \label{fig:3}
\end{figure}

\begin{figure}
  \centering
  \includegraphics[width=\columnwidth]{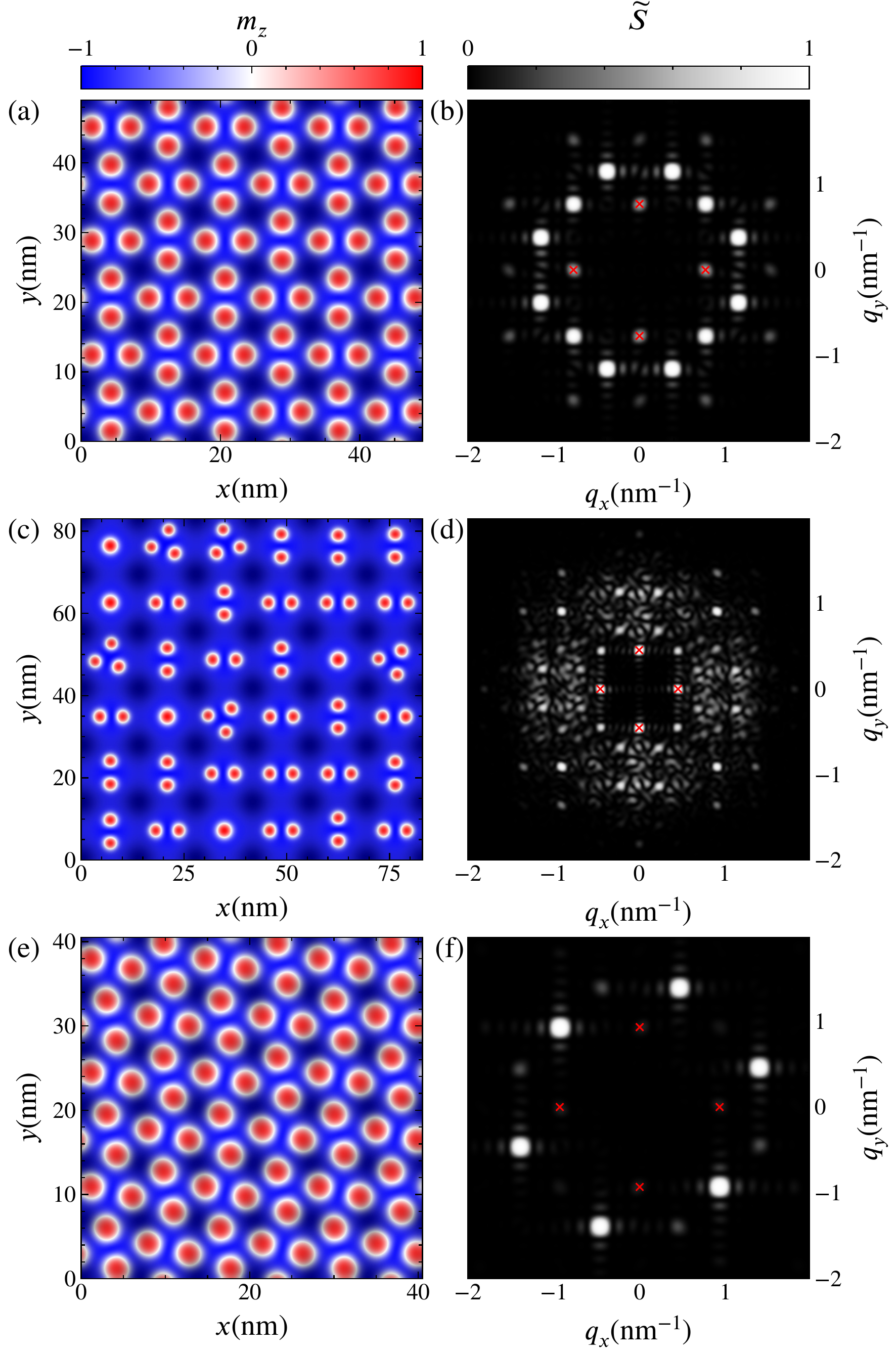}
  \caption{
    Configurations of the $N_\mathrm{sk}/N_m=2$ skyrmion
    lattice for selected states from the phase diagram in
    Fig.~\ref{fig:2}. (a, c, e) Real space images.
    (b, d, f) Reciprocal space images.
    (a, b) AFM phase at $\mu H=0.25D^2/J$ and $L=49$ nm.
    (c, d) Disordered state at $\mu H=0.58D^2/J$ and $L=83$ nm.
    (e, f) Floating triangular lattice at $\mu H=0$ and $L=40.5$ nm.
  }
  \label{fig:4}
\end{figure}

We first consider the ordering for a filling of $f = 2.0$.
In Fig.~\ref{fig:2} we summarize the different states
that occur in a phase diagram as a function of
applied field $\mu H$ versus system size $L$.
For small $L = 32$ nm and larger $\mu H$, half the skyrmions
annihilate, and the system forms a square lattice with $f = 1.0$ as shown in Fig.~\ref{fig:3}(a)
for $\mu H=0.33D^2/J$ and $L=32$ nm.
In Fig.~\ref{fig:3}(b) we plot
the corresponding reciprocal space
image, where the red $\times$ marks indicate the locations of the
trivial peaks from the square substrate lattice.
The superlattice state from
Fig.~\ref{fig:2} is illustrated
in Fig.~\ref{fig:3}(c) at $\mu H=0.58 D^2/J$ and $L=66$ nm.
In this case, there is no skyrmion
annihilation so that we still have $f = 2.0$,
but the substrate minima are filled with an alternating pattern
of single skyrmions and triple skyrmions.
This state can be
regarded as a checkerboard pattern of coexisting
monomers and trimers.
The skyrmion sizes are not constant; the skyrmion monomers are the largest,
while the trimers are assembled from
one larger skyrmion and two smaller skyrmions.
The trimers exhibit orientational ordering and are all aligned
in the same direction.
The corresponding reciprocal space plot
in Fig.~\ref{fig:3}(d) has a much more complicated pattern due
to the superlattice ordering.
As shown in
Fig.~\ref{fig:2}(e, f) at $\mu H=0.58D^2/J$ and $L=108.5$ nm,
we also find a dimer state in which each potential minimum
captures two skyrmions and neighboring dimers are perpendicular
to each other in what we call an
antiferromagnetic (AFM) dimer state.
The AFM dimer arrangement is the same as the AFM colloidal dimer state
observed for colloids on a square substrate
\cite{Reichhardt02, Agra04},
indicating that for these parameters,
the skyrmion dimers act like an object with a quadrupole moment,
similar to the behavior of the colloidal system.
The phase diagram of Fig.~\ref{fig:2} indicates that
the AFM dimer state appears in two windows, with one occurring
at $32~\mathrm{nm} < L < 74$ nm and smaller $\mu H$,
and the other at higher $\mu H$ and larger $L$.
In Fig.~\ref{fig:4}(a, b) we show the real and reciprocal space configurations
for the small $L$ AFM phase at
$\mu H=0.25D^2/J$ and $L=49$ nm.
The antiferromagnetic ordering persists,
but the individual skyrmions are much larger compared
to the system with larger $L$.

For higher $\mu H$ and larger $L$, we find a tilted dimer lattice state,
as illustrated in
Fig.~\ref{fig:3}(g, h) at $\mu H=1D^2/J$ and $L=100$ nm.
Here, the dimers are aligned along the $45^\circ$
diagonals, and adjacent dimers adopt opposite tilt angles.
Interestingly, the tilted dimer state was not
observed in previous work on colloids at fillings of $f=2.0$
for a square substrate \cite{Reichhardt02, Agra04, ElShawish08, ElShawish11}.
This may indicate that
the interactions between adjacent skyrmions differs from
the effective quadrupole interactions between Yukawa point
particles found in the colloid system
\cite{Agra04}.

At lower $\mu H$ and higher $L$, the skyrmions become elongated,
but the system can still maintain long range
positional and orientational order to form what we call a stretched phase,
as shown in real and reciprocal space
in Fig.~\ref{fig:3}(i, j) at $\mu H=0.25D^2/J$ and $L=117$ nm.
The ordering is the same
as in the ordered tilted dimer state,
but the skyrmion in each dimer has stretched out perpendicular to the
dimer axis.

The phase diagram of Fig.~\ref{fig:2} indicates that disordered states
appear along the boundary of the superlattice state.
These disordered states typically
contain skyrmions of different sizes,
as shown in Fig.~\ref{fig:4}(c, d) at $\mu H=0.58D^2/J$ and $L=83$ nm.
Most of the minima contain dimers but there are some monomers and trimers
present; in addition, the dimers do not have a consistent orientational
ordering.
The corresponding reciprocal space
image in Fig.~\ref{fig:4}(d)
shows the four trivial peaks associated with the substrate,
but the remaining peaks
are smeared out, indicating a lack of long range orientational order.
The disordered phases are likely due to frustration,
since there are multiple ways
in which the system can reach a low energy configuration.

\begin{figure}
  \centering
  \includegraphics[width=\columnwidth]{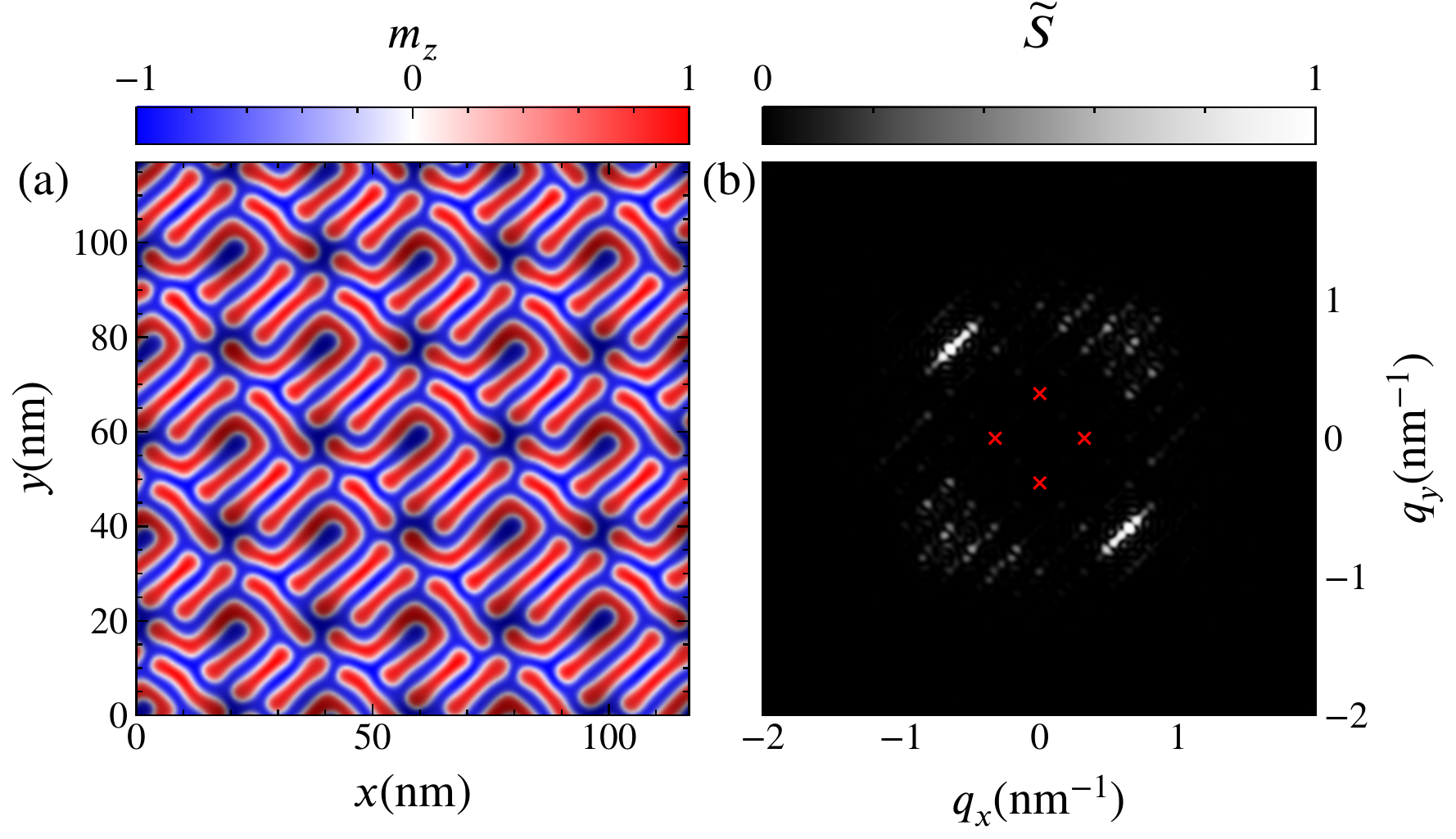}
  \caption{
    Configurations of the $N_\mathrm{sk}/N_m=2$ skyrmion
    lattice for the ordered stripe phase at $\mu H=0$ and $L=117$
    from the phase diagram in Fig.~\ref{fig:2}. (a) Real space image.
    (b) Reciprocal space image.
  }
  \label{fig:5}
\end{figure}

There is a region in Fig.~\ref{fig:2}
where triangular lattice ordering extends
all the way down to $\mu H=0$,
as shown in Fig.~\ref{fig:4}(e, f) at $\mu H=0$ and $L=40.5$ nm.
No skyrmion annihilation occurs, so the state retains
the $f = 2.0$ filling.
The skyrmion-skyrmion interactions become
strong enough to dominate over the interaction with the square
substrate, and a triangular skyrmion lattice emerges that is
floating over the substrate.
In previous work with hard-sphere colloids on a square substrate,
transitions from a square to a triangular lattice
appeared as the packing density increased \cite{Neuhaus13}.
For the skyrmions,
the triangular lattice emerges for smaller substrate
spacings, so that the system can be viewed as being more strongly packed.
At $\mu H=0$ in equilibrium, a skyrmion state is not
stable;
however, experimentally,
it is possible to supercool skyrmions from a finite field to
$\mu H=0$ in order to achieve a metastable skyrmion state.
In our system, we seed the skyrmions during
the initialization, and therefore
they can remain in a metastable
triangular or square lattice even at zero magnetic field.
The stretched state that appears for larger substrate lattice spacings
at $\mu H=0$ has a more complex geometry
than the stretched dimer
configuration shown in in Fig.~\ref{fig:3}(i, j),
but it can still have long-range order,
as indicated in Fig.~\ref{fig:5}(a, b) at
$\mu H=0$ and $L=117$ nm,
where there is a repeating stripe pattern that is aligned with the
underlying square substrate.

\section{$f= 3.0$}

\begin{figure}
  \centering
  \includegraphics[width=\columnwidth]{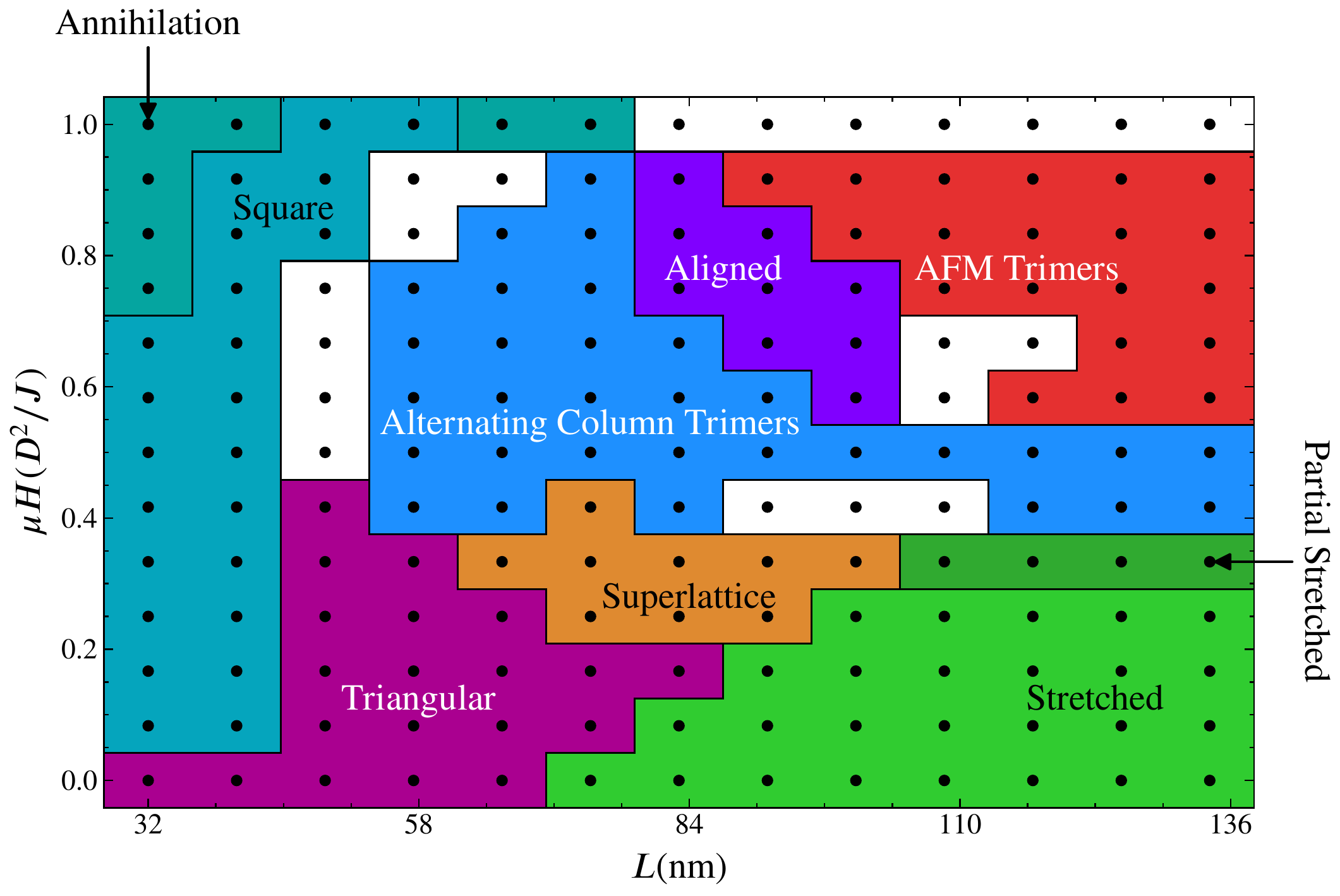}
  \caption{
    Phase diagram indicating the skyrmion ordering
    as a function of magnetic
    field $\mu H$ vs system size $L$
    for a system with $N_\mathrm{sk}/N_m=3$ as the
    starting condition.
    Turquoise: all skyrmions annihilate.
    Teal: square lattice.
    Magenta: triangular lattice.
    Orange: superlattice.
    Blue: alternating column trimers.
    Dark green: partially stretched state.
    Light green: stretched state.
    Purple: aligned state where all trimers are oriented in the same
    direction.
    Red:
    AFM trimer state.
    In white regions, the system is either disordered or does not have
    a well defined ordering.
  }
  \label{fig:6}
\end{figure}

\begin{figure}
  \centering
  \includegraphics[width=\columnwidth]{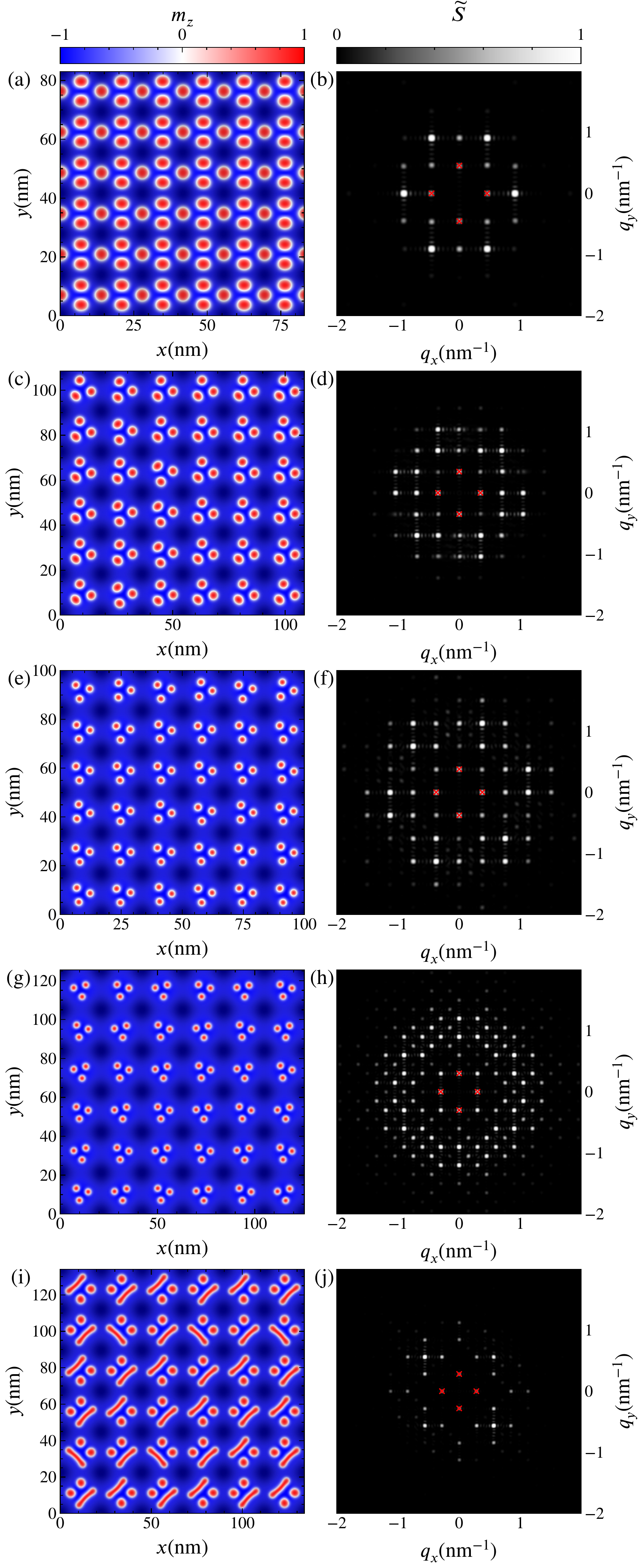}
  \caption{Configurations of the $N_\mathrm{sk}/N_m=3$ skyrmion
    lattice for selected states from
    the phase diagram in Fig.~\ref{fig:6}.
    (a, c, e, g, i) Real space images.
    (b, d, f, h, j) Reciprocal space images.
    (a, b) Superlattice at $\mu H=0.25D^2/J$ and $L=83$ nm.
    (c, d) Alternating column trimers at $\mu H=0.42D^2/J$ and $L=108.5$ nm.
    (e, f) Aligned dimers at $\mu H=0.67D^2/J$ and $L=100$ nm.
    (g, h) AFM dimers at $\mu H=0.67D^2/J$ and $L=125.5$ nm.
    (i, j) Stretched state at $\mu H=0.33D^2/J$ and $L=134$ nm.
  }
  \label{fig:7}
\end{figure}

We next consider the ordering for three skyrmions per substrate or
$f = 3.0$. A phase diagram as a function of $\mu H$ versus $L$ in
Fig.~\ref{fig:6} summarizes the states we observe, while images of some
of the phases in real and reciprocal space appear in
Fig.~\ref{fig:7}.
For small $L$ and high $\mu H=0$
in Fig.~\ref{fig:6},
all of the skyrmions annihilate.
There is an extended range of small
$L$ values at $\mu H=0$ for which
two thirds of the skyrmions annihilate
in order to produce a $f = 1.0$ square lattice.
As $L$ increases,
a region of low field triangular lattice appears, similar to what was
described above for the dimer system.
For fields near
$\mu H=0.3D^2/J$ we find a superlattice
ordering of the type
illustrated in Fig.~\ref{fig:7}(a, b) at
$\mu H=0.25D^2/J$  and $L=83$ nm.
We observe several distinct trimer phases, including
the alternating column phase shown in
Fig.~\ref{fig:7}(c, d) at $\mu H=0.42D^2/J$ and $L=108.5$ nm.
In this case, each substrate minimum
captures three skyrmions, and all of the resulting trimers in a
given column are tilted at the same angle, but there is a slight
variation of the angle from column to column. The
individual skyrmions adopt a combination of circular
and slightly elliptical shapes.
There are also regions of aligned trimers, as illustrated
in Fig.~\ref{fig:7}(e, f) at $\mu H=0.67D^2/J$ and $L=100$ nm,
where all of the trimers are aligned in the same direction and
the ordering is of ferromagnetic type.

The AFM trimer state is shown
in Fig.~\ref{fig:7}(g, h) at
$\mu H=0.67D^2/J$ and $L=125.5$ nm, and contains
well-defined trimers that form a bipartite lattice with two tilt
directions, with adjacent trimers having opposite tilts.
For colloids at $f = 3.0$ filling, simulations
produced
column ordered states in which the orientation of the trimers
was reversed
from one column to the next
\cite{Reichhardt02}.
Although this has the same bipartite column structure as
the AFM trimer state we observe for the skyrmions,
the difference in tilt angle is much more pronounced in the colloidal
system, and thus there is not an exact match between the
skyrmion and colloidal system in this state.

The aligned trimer and AFM trimer states
we find for skyrmions
were not observed in previous work in colloidal systems;
however, an aligned trimer state
appeared for colloids on a hexagonal lattice \cite{Reichhardt02, Brunner02}.
Thus, even when the skyrmions do
not stretch,
they can support more phases than a purely particle-based model.
This may be due to the ability of the skyrmion to distort even by
small amounts away from a completely circular shape.
Some other evidence for this is that skyrmions in liquid crystals interacting
with a square array of obstacles at a filling of
$f=3$,
when treated with a model in
which the liquid crystal skyrmions are able to distort \cite{Duzgun20},
produce the same tilted alternating trimer state that we find
at $f=3$.

\begin{figure}
  \centering
  \includegraphics[width=\columnwidth]{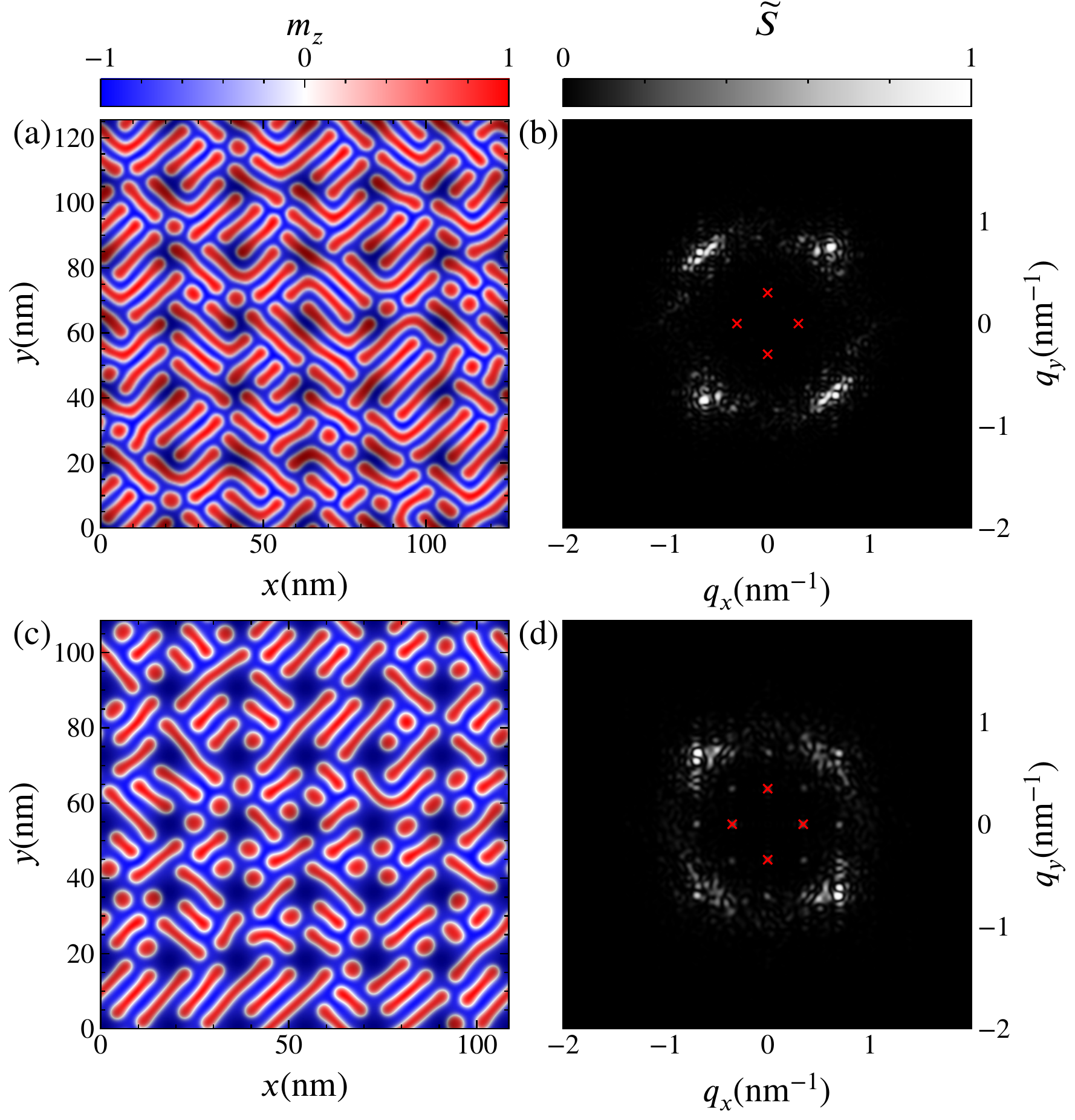}
  \caption{
    Configurations of the $N_\mathrm{sk}/N_m=3$ skyrmion
    lattice for selected states from the phase
    diagram in Fig.~\ref{fig:6}. (a, c) Real space images.
    (b, d) Reciprocal space images.
    (a, b) Disordered stretched state at $\mu H=0.0$ and $L=125.5$ nm.
    (c, d) Disordered stretched state at $\mu H=0.17$ and $L=108.5$ nm.
  }
  \label{fig:8}
\end{figure}

At lower fields we observe
a partially stretched state in which stretched and circular
skyrmions coexist,
as shown in Fig.~\ref{fig:7}(i, j) at $\mu H=0.33D^2/J$ and $L=134$ nm.
There is one stretched skyrmion and two circular skyrmions
in each minimum,
with the orientation of the alternating pattern flipping
from one column to the next.
As the field is lowered, the skyrmions become more elongated,
and the stretched state loses its ordering.
In Fig.~\ref{fig:8}(a, b), the $f=3.0$
$\mu H=0.0$ and $L=125.5$ nm stretched state  is composed
of stretched skyrmions
and a small number of circular skyrmions,
but there is no long range order.
A similar state appears
in Fig.~\ref{fig:8}(c, d)
at $\mu H=0.17$ and $L=108.5$ nm.
Unlike the $f = 2.0$ states, the stretched states for
the $f=3.0$ trimer fillings are not ordered.

\section{$f=3/2$}

\begin{figure}
  \centering
  \includegraphics[width=\columnwidth]{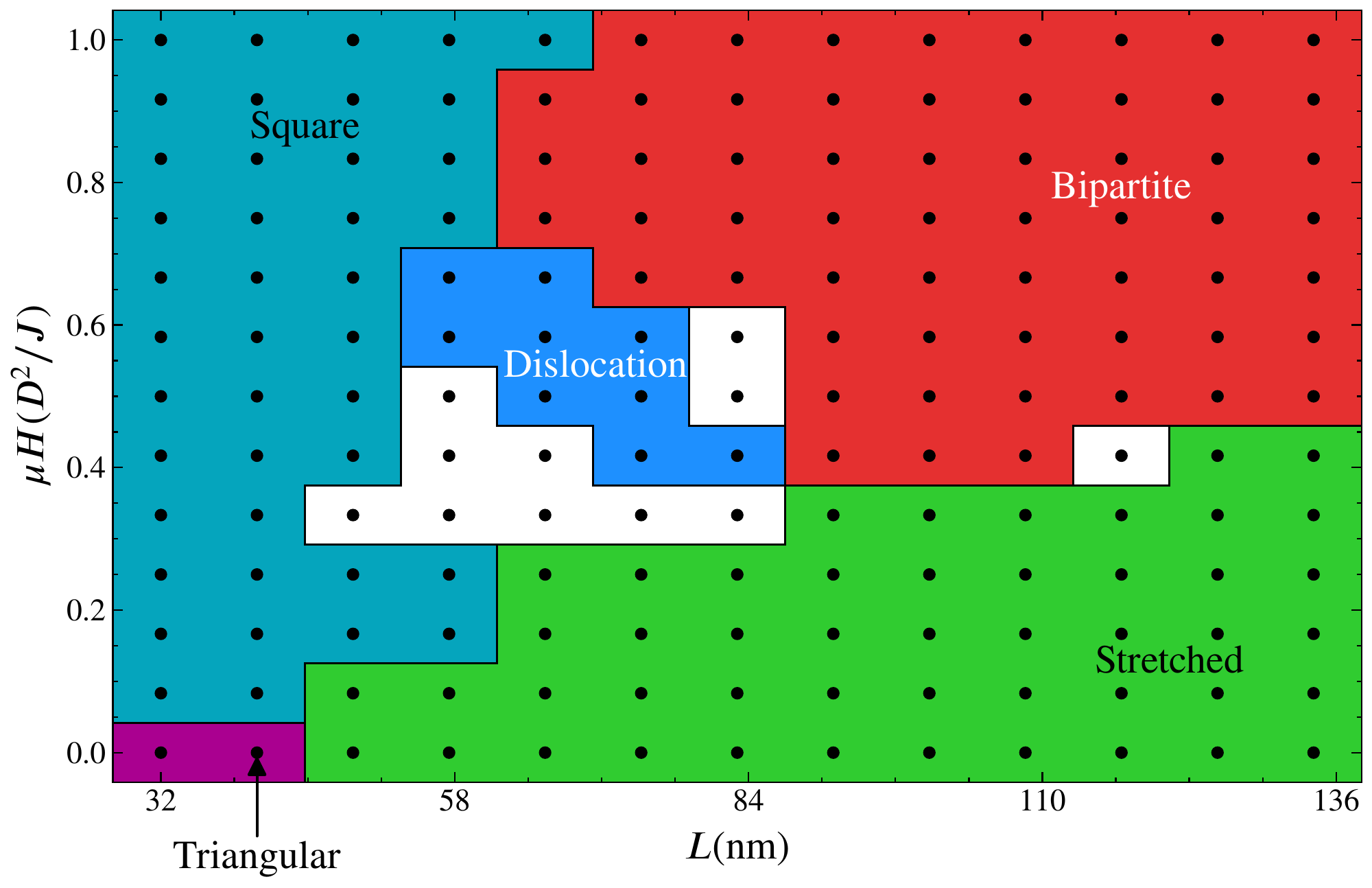}
  \caption{Phase diagram indicating the skyrmion ordering as
    a function of magnetic field $\mu H$ vs system size $L$
    for a system with $N_\mathrm{sk}/N_m=3/2$ as the
    starting condition.
    Teal: square lattice.
    Magenta: triangular lattice.
    Blue: dislocation state.
    Red: bipartite lattice.
    Green: stretched state.
    In white regions, the system is either disordered or does not
    have a well defined ordering.
  }
  \label{fig:9}
\end{figure}

\begin{figure}
  \centering
  \includegraphics[width=\columnwidth]{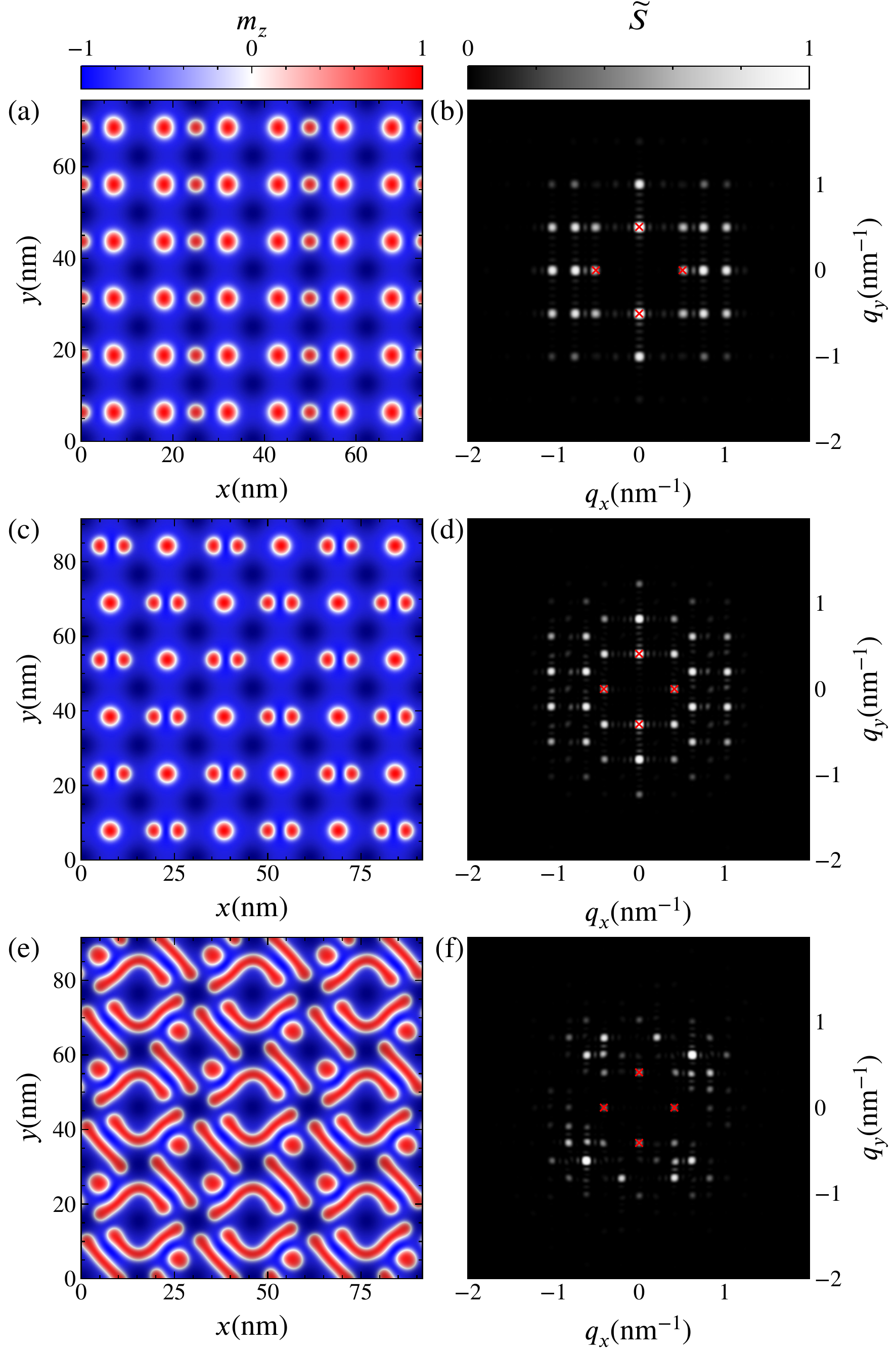}
  \caption{Configurations of the $N_\mathrm{sk}/N_m=3/2$ skyrmion
    lattice for selected states from the phase diagram in Fig.~\ref{fig:9}.
    (a, c, e) Real space images.
    (b, d, f) Reciprocal space images.
    (a, b) Dislocation state at $\mu H=0.42D^2/J$ and $L=74.5$ nm.
    (c, d) Bipartite lattice at $\mu H=0.42D^2/J$ and $L=91.5$ nm.
    (e, f) Stretched state at $\mu H=0.17D^2/J$ and $L=91.5$ nm.
  }
  \label{fig:10}
\end{figure}

We next consider the ordering at $f = 3/2$.
The different orderings are highlighted
in the phase diagram as
a function of $\mu H$ versus $L$ in Fig.~\ref{fig:9}.
For $L < 58$ nm, there is an extended region where a
portion of the skyrmions annihilate and a square lattice
with $f = 1.0$ appears.
For $\mu H=0.0$ and for $L < 45$ nm, there
is a narrow window of triangular lattice ordering.
A large region of
stretched state appears for
$L > 45$ nm and $\mu H < 0.4D^2/J$.
We also find what we call a dislocation phase where the system
forms a square commensurate lattice that has,
in every other column, a smaller skyrmion located on the
saddle point, as illustrated
in Fig.~\ref{fig:10}(a, b) at $\mu H=0.42D^2/J$ and $L=74.5$ nm.
Placing the additional skyrmion on the saddle point costs energy,
and as a result the saddle point skyrmions are reduced in size.
We observe an extended region of bipartite order where the system forms
a checkerboard
state of aligned dimers
alternating with monomers, as shown in Fig.~\ref{fig:10}(c, d)
at $\mu H=0.42D^2/J$ and $L=91.5$ nm.
Here, the dimers are aligned along the $x$ direction,
and the skyrmions in the dimers are smaller than the monomer skyrmions.
In addition to disordered stretched states, we find stretched
phases that can have
complex long-range ordering, such as the example plotted
in Fig.~\ref{fig:10}(e, f) at $\mu H=0.17D^2/J$ and $L=91.5$ nm.
In some regions of the phase diagram, no long range order is present.

\section{$f= 5/2$}
\begin{figure}
  \centering
  \includegraphics[width=\columnwidth]{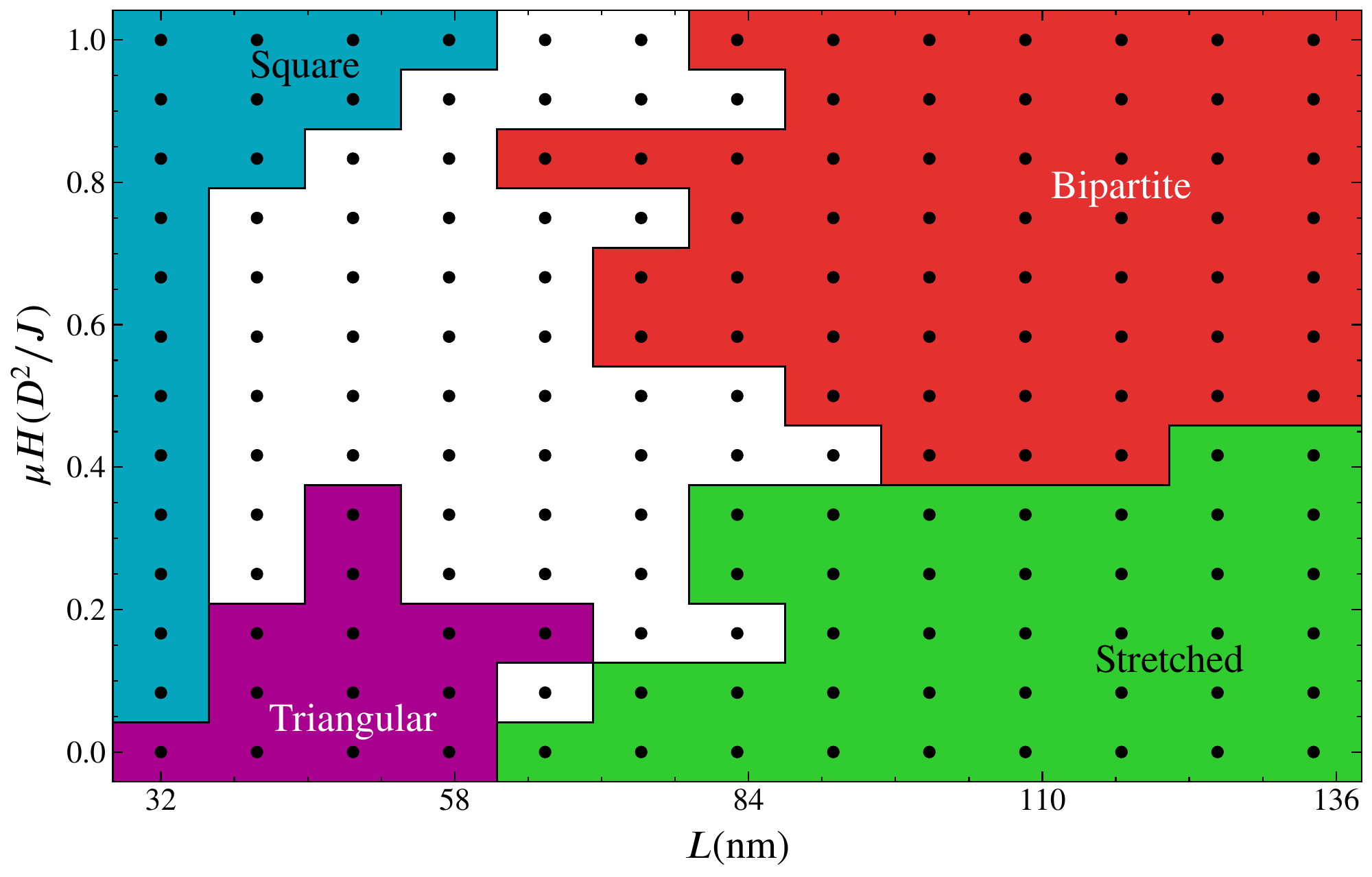}
  \caption{Phase diagram indicating the skyrmion ordering as a
    function of magnetic field $\mu H$ vs system size $L$
    for a system with $N_\mathrm{sk}/N_m=5/2$ as the
    starting condition.
    Teal: square lattice.
    Purple: triangular lattice.
    Red: bipartite lattice.
    Green: stretched state.
    In white regions, the system is either disordered or does
    not have a well defined ordering.
  }
  \label{fig:11}
\end{figure}

\begin{figure}
  \centering
  \includegraphics[width=\columnwidth]{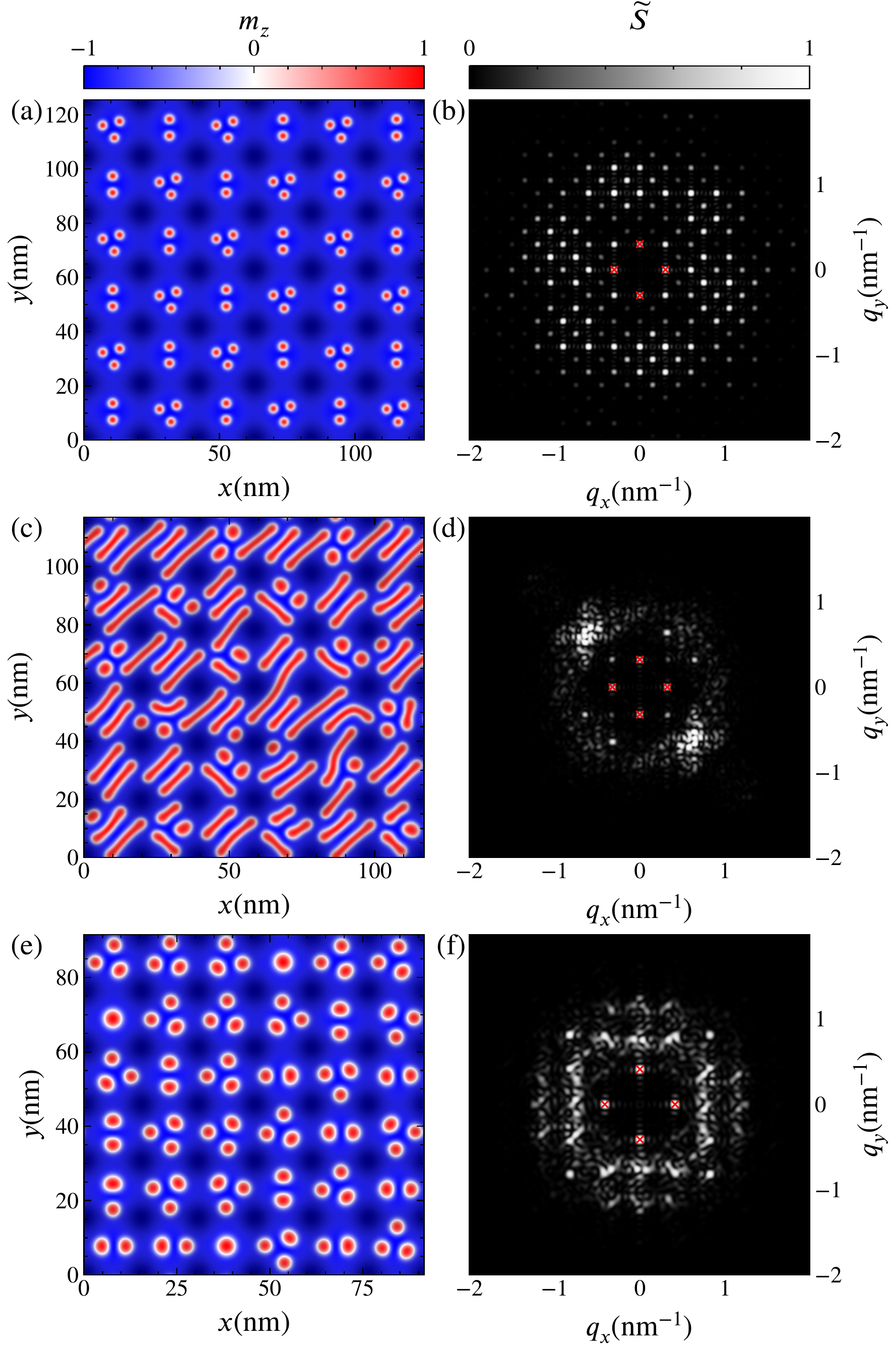}
  \caption{Configurations of the $N_\mathrm{sk}/N_m=5/2$ skyrmion
    lattice for selected states from
    the phase diagram in Fig.~\ref{fig:11}.
    (a, c, e) Real space images.
    (b, d, f) Reciprocal space images.
    (a, b) Bipartite state at $\mu H=0.67D^2/J$ and $L=125.5$ nm.
    (c, d) Stretched state at $\mu H=0.25D^2/J$ and $L=117$ nm.
    (e, f) Disordered state at $\mu H=0.42D^2/J$ and $L=91.5$ nm.
  }
  \label{fig:12}
\end{figure}

In Fig.~\ref{fig:11} we
show a phase diagram as a function of $\mu H$ versus $L$ for systems
initialized with a filling of
$f = 5/2$.
There is a square lattice regime where a portion of the skyrmions annihilate,
as well as a window of triangular lattice for small $L$.
A bipartite phase consisting of a mixture of
trimers and dimers appears for larger fields and system sizes,
as shown in Fig.~\ref{fig:12}(a, b)
for $\mu H=0.67D^2/J$ and $L=125.5$ nm,
where the dimers are aligned along the $y$ direction and
two of the skyrmions in the trimer are parallel to the 45$^\circ$ direction.
The $f=5/2$
stretched phase is generally disordered and consists
of elongated skyrmions mixed with monomers,
as illustrated in Fig.~\ref{fig:12}(c, d)
at $\mu H=0.25D^2/J$ and $L=117$ nm.
We also find an extended region of disordered phases for this filling,
as shown in Fig.~\ref{fig:12}(e, f) at $\mu H=0.42D^2/J$ and $L=91.5$ nm.

\section{Metastable States}

The orderings described above were obtained as
described in Sec.~\ref{sec:2} by performing
numerical integration of Eq.~\ref{eq:2}
with thermal effects. If
the thermal effects are omitted, other orderings corresponding
to metastable configurations emerge. We next describe
a few of the metastable orderings we observe when integrating
Eq.~\ref{eq:2} without including thermal fluctuations.

\begin{figure}
  \centering
  \includegraphics[width=\columnwidth]{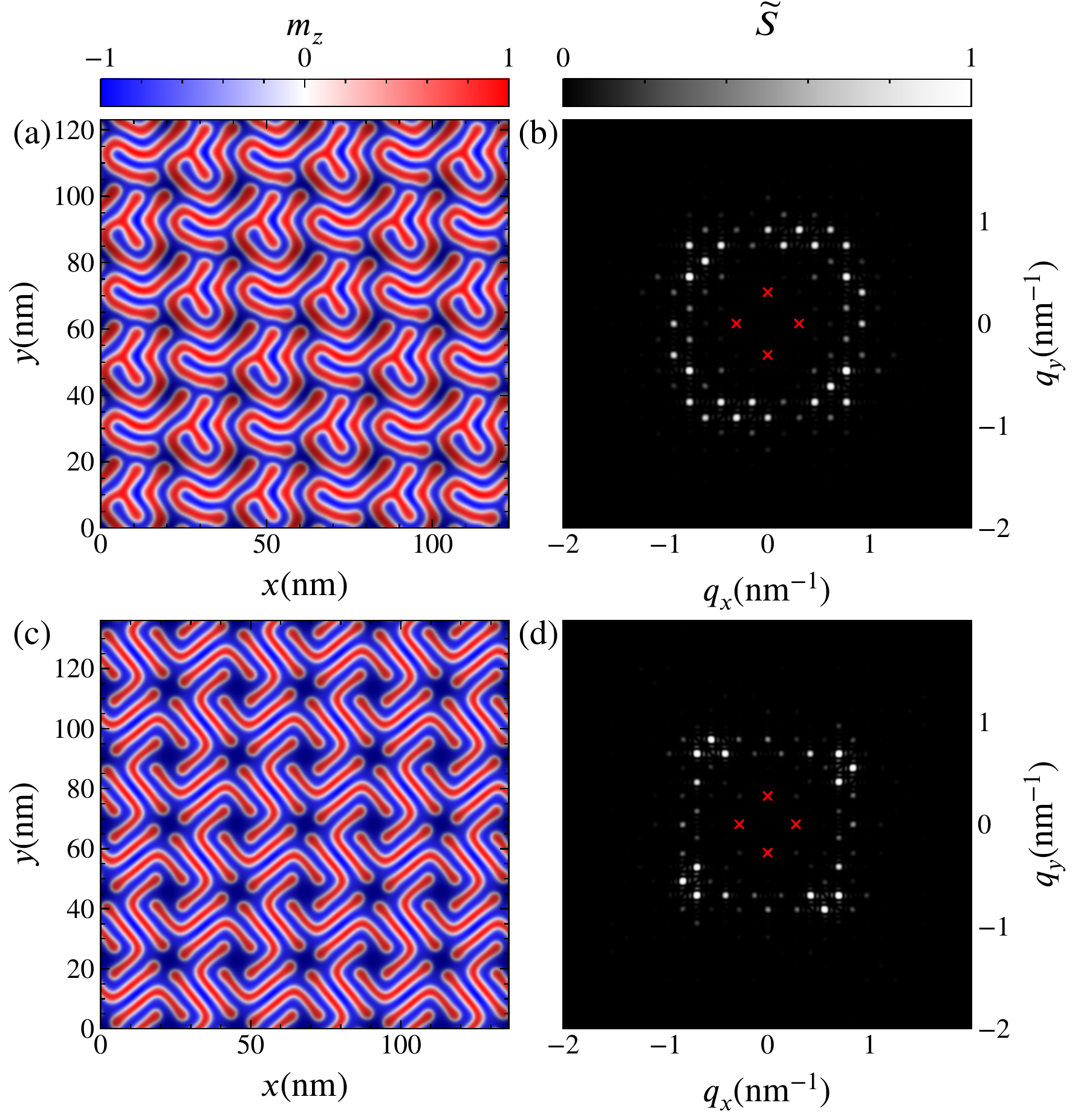}
  \caption{Configurations of the $N_\mathrm{sk}/N_m=2$ skyrmion lattice
    for selected
    metastable states in which no thermal fluctuations are included
    in the calculation.
    (a, c) Real space images. (b, d) Reciprocal space images.
    (a, b) An ordered stripe state at $\mu H=0.0$ and and $L = 123$ nm.
    (c, d) An ordered stripe state at $\mu H=0.15D^2/J$ and $L = 136$ nm.
  }
  \label{fig:13}
\end{figure}

\begin{figure}
  \centering
  \includegraphics[width=\columnwidth]{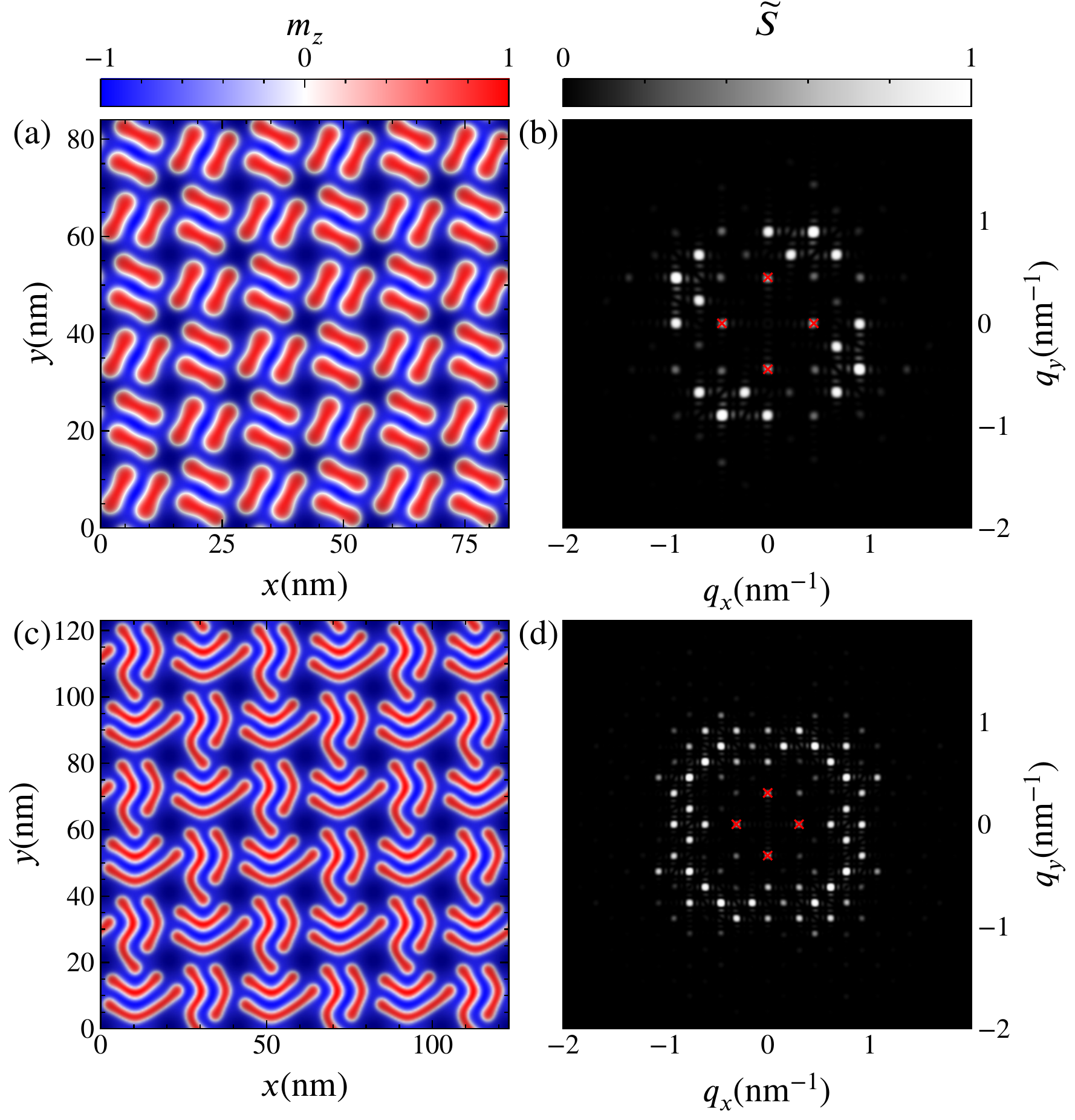}
  \caption{Configurations of the $N_\mathrm{sk}/N_m=2$ skyrmion
    lattice for selected metastable states in which no thermal
    fluctuations are included in the calculation.
    (a, c) Real space images. (b, d) Reciprocal space images.
    (a, b) A stretched state with alternating tilt
    at $\mu H=0.1D^2/J$ and $L = 84$ nm.
    (c, d) A stretched state with partial antiferromagnetic
    ordering at $\mu H=0.2D^2/J$ and $L = 123$ nm.
  }
  \label{fig:14}
\end{figure}

A number of the previously described phases at larger $L$
remain robust in the absence of thermal fluctuations,
but some additional stripe orderings appear.
For example, in Fig.~\ref{fig:13}(a, b) at $\mu H=0.0$ and $L = 123$ nm,
a complex stripe state with a periodic structure forms.
In Fig.~\ref{fig:13}(c, d),
a complex pattern of stretched out dimers with periodic chiral ordering
appears at $\mu H=0.15D^2/J$ and $L = 136$ nm.
For $\mu H=0.1D^2/J$ and $L = 84$ nm,
we find a stretched state with alternating tilt, as shown in
Fig.~\ref{fig:14}(a, b).
Finally,
in Fig.~\ref{fig:14}(c, d) we illustrate
a stretched state with partial antiferromagnetic ordering
at $\mu H=0.2D^2/J$ and $L = 123$ nm.

\section{Discussion}

The orientationally ordered $n$-mer
states we observe should also have interesting dynamics, such
as sliding under dc driving or pattern switching for weaker drives.
Some aspects of the dynamics
have already been studied for dc and ac driving of a
small subset of the states
described here \cite{Souza24A, Souza25a};
however, we expect that a variety of other kinds of dynamics will
emerge at other
ordered or partially ordered states.
Under a dc drive, the antiferromagnetic state could undergo
a transition to a ferromagnetic polarized state in which
the dimers or trimers are aligned with the drive.
It would also be interesting to explore
the frustrated state
in more detail to see if
this system is more similar
to spin ices with strong degeneracy,
or is instead closer to a glass that becomes
kinetically trapped before reaching a unique ground state.
In most spin ice systems,
frustration arises as a result of the flipping or rotation of spins;
however, for the skyrmions there
are additional ways to produce frustration through changes in
skyrmion shape or size.

Another question is whether increasing the thermal effects could lead to
thermally induced transitions.
In colloidal spin ice, as the thermal fluctuations become more
important, a transition can occur to a phase in which
the dimers and trimers have lost their orientational ordering, while
at high temperatures the particles can hop between substrate minima
\cite{Brunner02, Reichhardt02}.
A future study of increasing temperature
could investigate whether two-step melting
similar to what is seen for the colloids can occur for
skyrmions, or whether the skyrmions instead undergo
a single disordering transition to a strongly fluctuating state.
The disordered states we observe could be more susceptible
than the ordered states to thermal motion
if the barrier separating different configurations is low,
as is the case for frustrated states.

Our results could be tested experimentally
using skyrmions that have been coupled to a periodic array of magnetic dots,
periodic thickness modulations, or an ion patterned sample.
It could also be possible for skyrmions to be trapped by an optical lattice.
Beyond skyrmions, the effects we describe are relevant to
other textures such as skyrmions in oblique crystals,
skyrmioniums, antiskyrmions, or biskyrmions \cite{Gobel21, Souza25b}.

\section{Summary}
We have examined skyrmion ordering on a square substrate created by a
periodic variation of the anisotropy
for fillings of $f=1.5$, 2.0, 2.5, and $3.0$ skyrmions
per substrate minimum for varied magnetic field and substrate lattice constant.
We observe several states in which the skyrmions
form ordered dimer or trimer lattices that have both positional
and orientational ordering,
similar to what has been seen for
colloidal molecular crystal states.
The ability of the skyrmions to annihilate or deform
produces a number of states not observed
in particle-based systems,
such as striped phases with long-range order
or stretched phases in which all or a portion of
the skyrmions become stretched or heavily distort into a stripe or
meron configuration.
The stretched phases are ordered for some fillings and disordered
for others.
There are also certain parameters for which the skyrmions do not
exhibit long-range order.
For a filling of two skyrmions per substrate minimum,
we find a rich variety of phases including
an antiferromagnetically ordered dimer state similar to that observed
at the same filling for colloids on square substrates.
Phases that appear only for the skyrmion system and not in
particle-based systems include superlattice ordering
where skyrmions of different sizes are present,
stretched dimer states, and a tilted dimer state.
For some parameters,
a portion of the skyrmions annihilate so that the
system can lower its filling fraction to form a commensurate square lattice.
We also find a floating solid phase in which
a triangular skyrmion lattice forms
that interacts only very weakly with the substrate.
For a filling of $f=3.0$, we
observe alternating ordered columnar states, superlattices, aligned
antiferromagnetic trimer states, and ordered and disordered stretched states.
For fillings of $f=1.5$ and $f=2.5$,
we find several orientationally and positionally
ordered bipartite lattices containing mixtures
of monomers, dimers, or trimers.
Our results provide a method for creating
different types of periodic skyrmion structures.
The states we observe could also appear in
other deformable particle-like systems
interacting with a square substrate, such
as charged droplets, other magnetic textures, and liquid crystal skyrmions.

\section*{Acknowledgments}
This work was supported by the US Department of Energy through the Los Alamos National Laboratory. Los
Alamos National Laboratory is operated by Triad National Security, LLC, for the National Nuclear Security
Administration of the U. S. Department of Energy (Contract No. 892333218NCA000001).
J. C. B. S and N. P. V. acknowledge funding from Fundação de Amparo à Pesquisa do Estado de São Paulo - FAPESP (Grants J. C. B. S 2023/17545-1 and 2022/14053-8, N. P. V. 2024/13248-5).
We would like to thank FAPESP for providing the computational resources used in this work (Grant: 2024/02941-1).

\bibliography{mybib}

\end{document}